\documentclass{ws-procs9x6}

\newcommand{\ee}{\mbox{$ \mathrm{e^+e^-}$}}
\newcommand{\cal}{\mbox{}}

\begin{document}
\thispagestyle{empty}

\def\thefootnote{\fnsymbol{footnote}}       

\begin{center}
\mbox{ } 
\vspace{-3cm}

\end{center}
\begin{flushright}
\Large
\mbox{\hspace{10.2cm} hep-ph/0403087} \\
\mbox{\hspace{11.1cm} March 2004}
\end{flushright}
\begin{center}
\vskip 1.0cm
{\Huge\bf
Exploring Supersymmetry at a Future Global \boldmath$\rm e^+e^-$\unboldmath

\vskip 0.4cm
 Linear Collider}
\vskip 1cm

{\LARGE\bf Andr\'e Sopczak}\\
\smallskip
\Large Lancaster University

\vskip 2.5cm
\centerline{\Large \bf Abstract}
\end{center}

\vskip 0.5cm
\hspace*{-1cm}
\begin{picture}(0.001,0.001)(0,0)
\put(,0){
\begin{minipage}{13cm}
\Large
\renewcommand{\baselinestretch} {1.2}
This review illustrates by means of sample reactions the potential 
of a future global $\rm e^+e^-$ Linear Collider (LC) for precision 
measurements of Supersymmetric particles 
with emphasis on recent studies and addressing major research directions.
\renewcommand{\baselinestretch} {1.}

\normalsize
\vspace{3cm}
\begin{center}
{\sl \large
\uppercase{P}resented at {\it \uppercase{SUSY} 2003:
\uppercase{S}upersymmetry in the \uppercase{D}esert}\/, 
held at the \uppercase{U}niversity of \uppercase{A}rizona,
\uppercase{T}ucson, \uppercase{AZ}, \uppercase{J}une 5-10, 2003. \\
\uppercase{T}o appear in the \uppercase{P}roceedings.
\vspace{-6cm}
}
\end{center}
\end{minipage}
}
\end{picture}
\vfill



\newpage
\thispagestyle{empty}
\mbox{ }
\newpage
\setcounter{page}{1}

\title{Exploring Supersymmetry at a Future Global 
       $\rm e^+e^-$ Linear Collider}

\author{ANDRE SOPCZAK}

\address{Department of Physics \\
Lancaster University\\ 
Lancaster LA1 4YW, United Kingdom\\ 
E-mail: Andre.Sopczak@cern.ch}

\maketitle

\abstracts{
\vspace*{-0.8cm}
This review illustrates by means of sample reactions the potential 
of a future global $\rm e^+e^-$ Linear Collider (LC) for precision 
measurements of Supersymmetric particles 
with emphasis on recent studies and addressing major research directions.
\vspace*{-0.8cm}
}

\section{Introduction}
\vspace*{-0.2cm}

Many searches for Supersymmetric particles and stringent limits which guide
future searches have been set by the LEP experiments
in an almost background-free experimental environment.
Currently, the Tevatron $\rm p\bar p$ collider, taking data at $\sqrt s =2$~TeV,
gives important search sensitivity with larger kinematic threshold.
When the LHC $\rm pp$ collider starts to operate at $\sqrt s =14$~TeV 
the kinematic reach will be increased significantly.
About 13 years of $\rm e^+e^-$ Linear Collider SUSY studies have been
performed with the focus changing from discovery to high-precision analyses.  
Recent milestones are 
the TESLA Technical Design Report 2001, 
the Snowmass 2001 meeting, and
the Linear Collider Workshops in Korea (2002) and Amsterdam (2003).
This review focuses on new developments for a 500 GeV to 3~TeV LC 
and includes preliminary results.

\vspace*{-0.3cm}
\section{The Linear Collider Project}

\subsection{Accelerator}
\vspace*{-0.2cm}
Important accelerator parameters are summarized in 
Table~\ref{tab:acc}.\,A future LC is 
characterized by 
high luminosity,
tunable centre-of-mass energy,
low beamstrahlung,
beam polarization,
additional options for $\gamma\gamma$ and $\rm e^-e^-$ collisions.

\begin{table}[htp]
\vspace*{-0.2cm}
\tbl{Accelerator parameters.}
{\footnotesize
\begin{tabular}{@{}c|cc|cc|c@{}}
\hline
Parameter  &  \multicolumn{2}{|c|}{TESLA} 
                 &  \multicolumn{2}{|c|}{ NLC/JLC} & CLIC \\
\hline
$\sqrt s$ (GeV)  & 500 &800    & 500 &1000     & 3000  \\
Gradient (MV/m)  & 23  &35     & 48  &48       & 150   \\
${\cal L}_{\rm cur}$ ($10^{34}{\rm cm}^{-2}{\rm s}^{-1}$)
                 & 3.4 &5.8    & 2.0 &3.4      & 10    \\
${\cal L}_{\rm int}/10^7$s (fb$^{-1}$)
                 & 340 &580    & 200 &340      & 1000  \\
Beamstrahlung spread (\%)
                 & 3.2 &4.3    & 4.7 &10.2     & 31    \\
\hline
\end{tabular}\label{tab:acc} }

\vspace*{-0.2cm}
\end{table}

\vspace*{-0.2cm}
\subsection{Detector}
\vspace*{-0.2cm}
Extensive and increasing R\&D for all sub-detectors is being performed.
An example is the development of CCD vertex detectors in the 
LC Flavour Identification (LCFI) collaboration\cite{lcfi}.
\vspace*{-0.1cm}

\section{Scalar Top Simulation with c-Quark Tagging}
\vspace*{-0.2cm}

\begin{figure}[htb]
\vspace*{-1.7cm}
\includegraphics[scale=0.27]{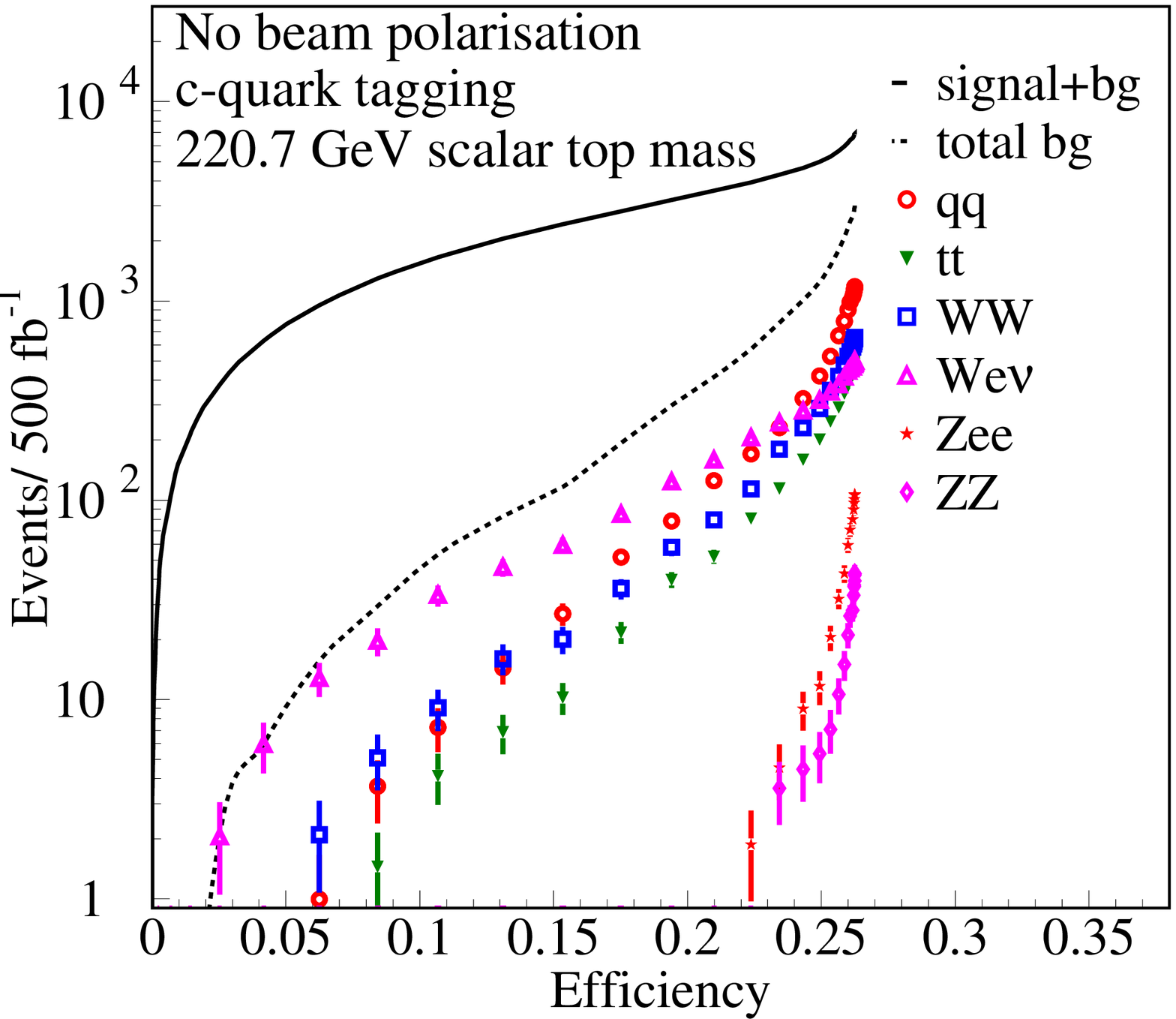}\hfill
\includegraphics[scale=0.3]{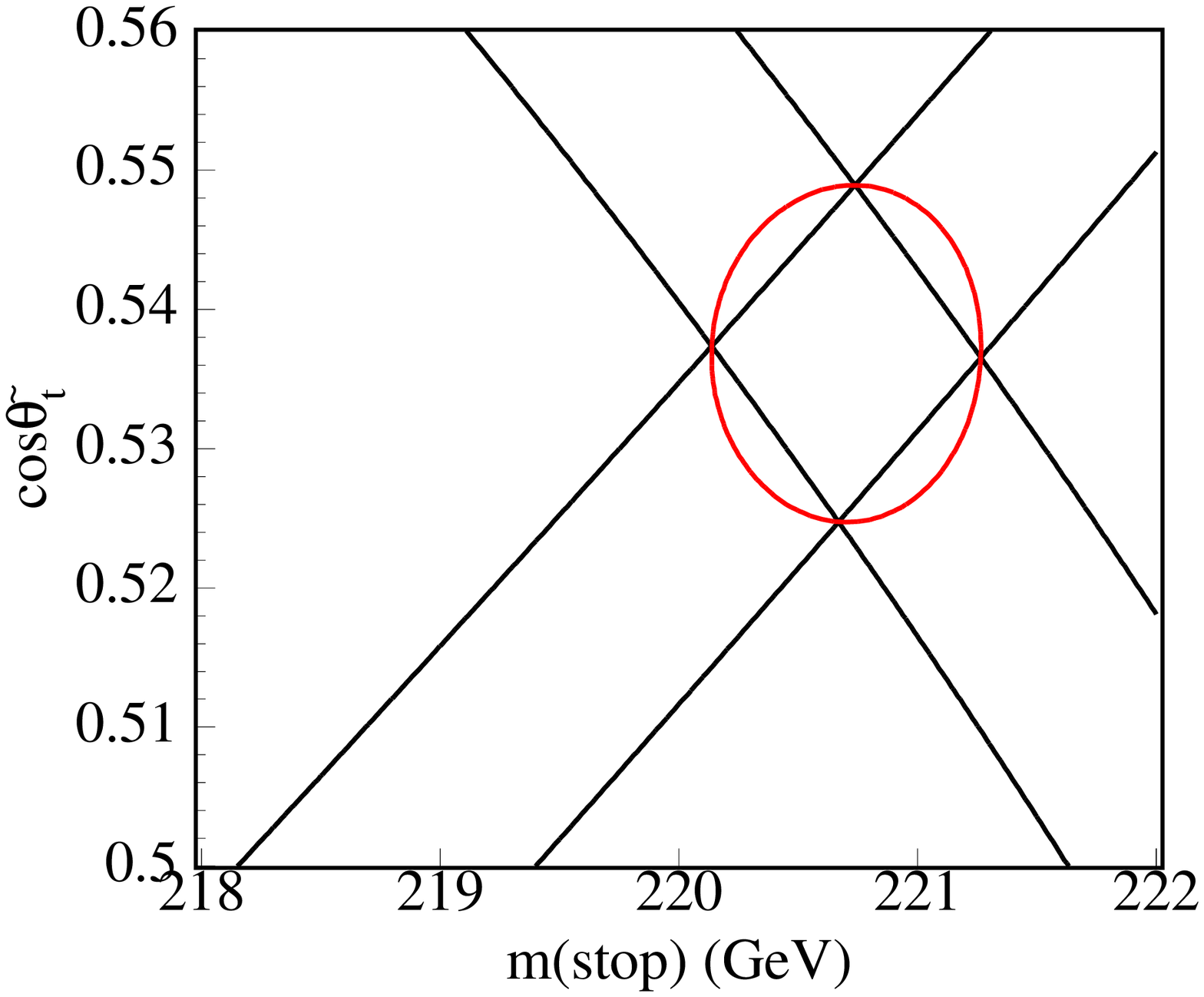}
\vspace*{-0.7cm}
\caption[]{\label{fig:massmixing}
Scalar top quark studies for SPS-5 parameters.
Left: expected number of signal and background events with c-quark tagging
for unpolarized beams. 
Right: sensitivity for mass and mixing angle from production cross section
precision determinations for 
$\sqrt{s}=500$~GeV and a total luminosity of $L=2\times500$~fb$^{-1}$
with ${\cal P}_{\rm e^-}=-0.80$, ${\cal P}_{\rm e^+}=0.60$ (left-polarization)
and
${\cal P}_{\rm e^-}=0.80$, ${\cal P}_{\rm e^+}=-0.60$ (right-polarization).
}
\vspace*{-0.4cm}
\end{figure}

In particular, hadronic background can be reduced with c-quark tagging
in the reaction 
$\rm \ee\rightarrow \tilde t_1\bar{\tilde t}_1 
        \rightarrow \tilde\chi^0_1c \tilde\chi^0_1\bar c$\cite{finch}.
The expected signal and background rates and resulting sensitivities for 
the scalar top mass and mixing angle determination
are shown in Fig.~\ref{fig:massmixing} for benchmark parameters SPS-5 
(mSUGRA)
$m_0=150$~GeV, $m_{1/2}=300$~GeV, $A_0=-1000$~GeV, 
$\tan\beta=5$, $\mu>0$, leading to
$m_{\rm \tilde t_1}=220.7\pm0.6$~GeV and 
$\cos\theta_{\rm \tilde t_1}=0.537\pm0.012$.

\vspace*{-0.35cm}
\section{Complex~Phases~on~Scalar~Top~and~Scalar~Bottom~Decays}
\vspace*{-0.2cm}

\begin{figure}[htb]
\vspace*{-0.2cm}
\includegraphics[scale=0.3]{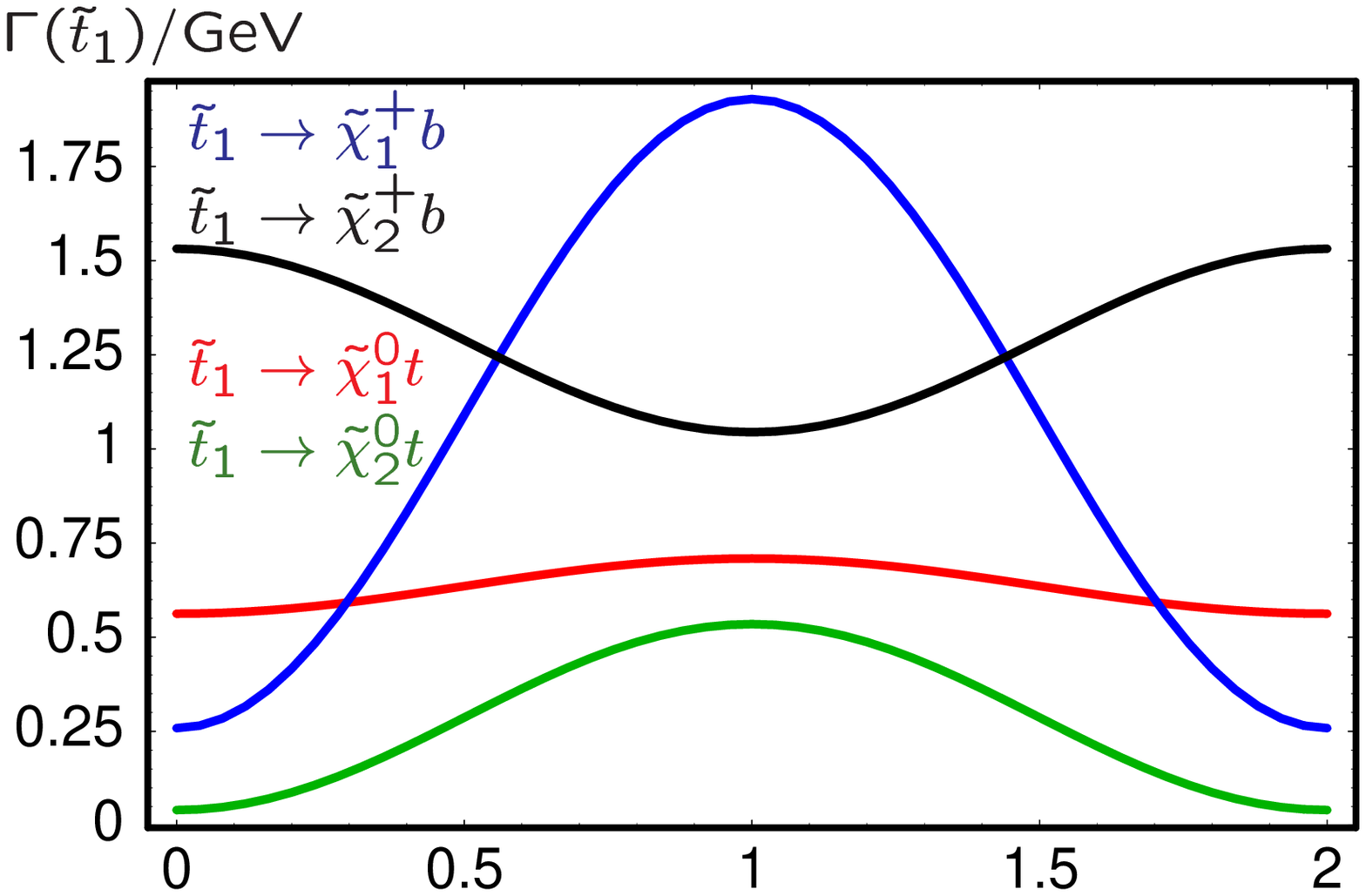}\hfill
\includegraphics[scale=0.3]{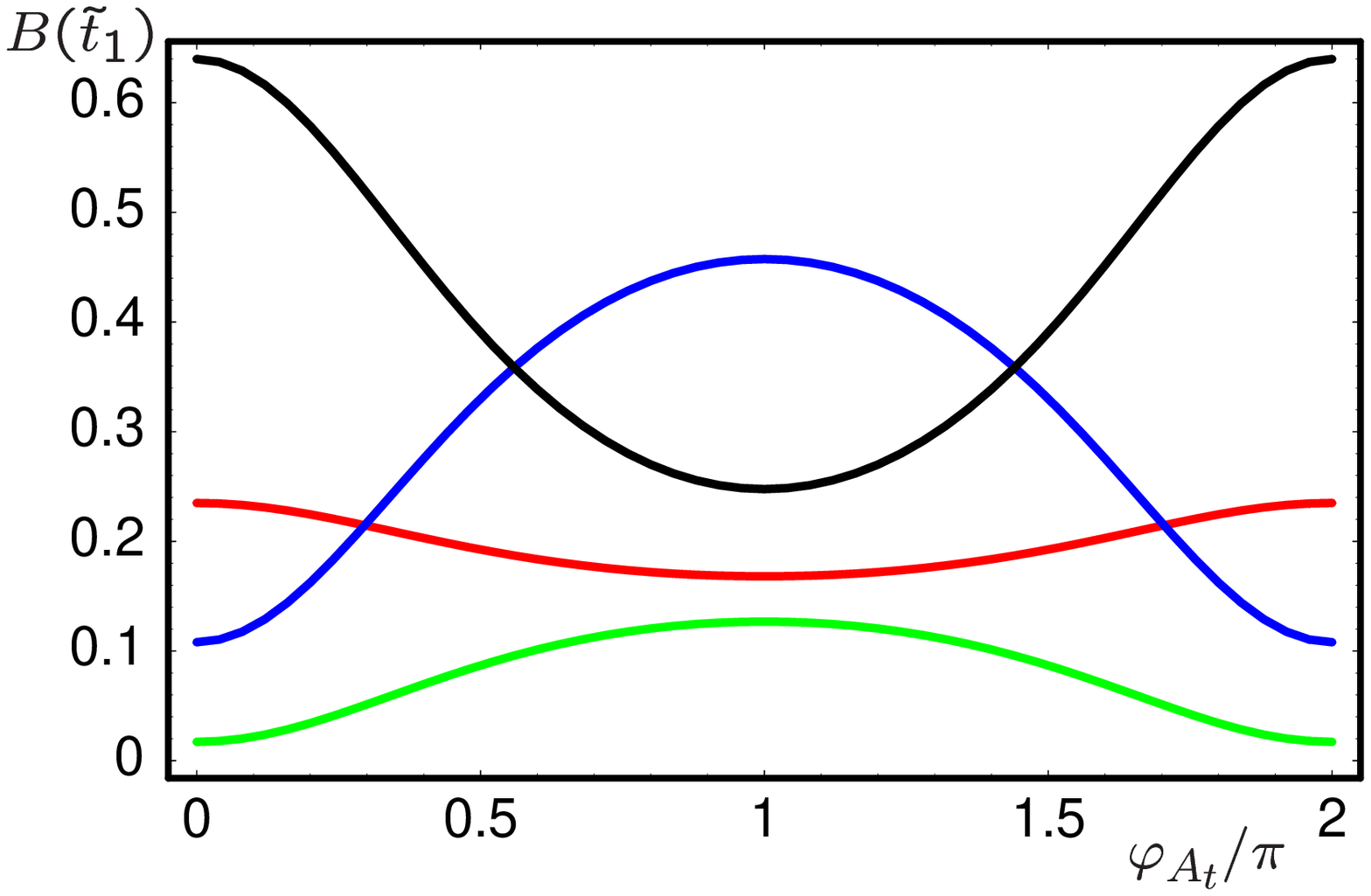}
\vspace*{-0.2cm}
\caption[]{\label{fig:complex}
Scalar top decay width and branching ratios as a function of the 
complex phase of the trilinear coupling of scalar fermion $A_f$.
}
\vspace*{-0.3cm}
\end{figure}

Complex phases in the Higgs potential and soft SUSY breaking terms
change the scalar top decay width and branching ratios\cite{bartl},
as shown in Fig.~\ref{fig:complex} 
for the example of a SPS-4 inspired scenario with 
$m_{\rm \tilde t_1}=531$~GeV.

\vspace*{-0.35cm}
\section{Scalar Muon Simulation}
\vspace*{-0.2cm}

The reaction
$\rm \ee\rightarrow\tilde \mu^+_{\rm R}\tilde\mu^-_{\rm R}
        \rightarrow\mu^+\tilde\chi^0_1\mu^-\tilde\chi^0_1$
has been re-investigated\cite{hanna}~for
benchmark\,parameters\,(SPS-1a)\,with\,$m_{\tilde\mu_{\rm R}}$=145.9\,GeV and
\mbox{$m_{\tilde\chi^0_1}$=100\,GeV.}
The isotropic scalar muon decay leads to a flat energy spectrum.
From a fit of this spectrum (end-point method)
the following precisions are obtained
\mbox{$m_{\tilde\mu_{\rm R}}=146.25\pm0.15$} GeV and $m_{\tilde\chi^0_1}=99.98\pm0.09$ GeV
for $\sqrt{s}=500$~GeV and $L=400$~fb$^{-1}$.

\vspace*{-0.2cm}
\section{$\tan\beta$ Sensitivity from Scalar Taus}
\vspace*{-0.2cm}

\begin{figure}[htb]
\vspace*{-0.4cm}
\includegraphics[scale=0.3]{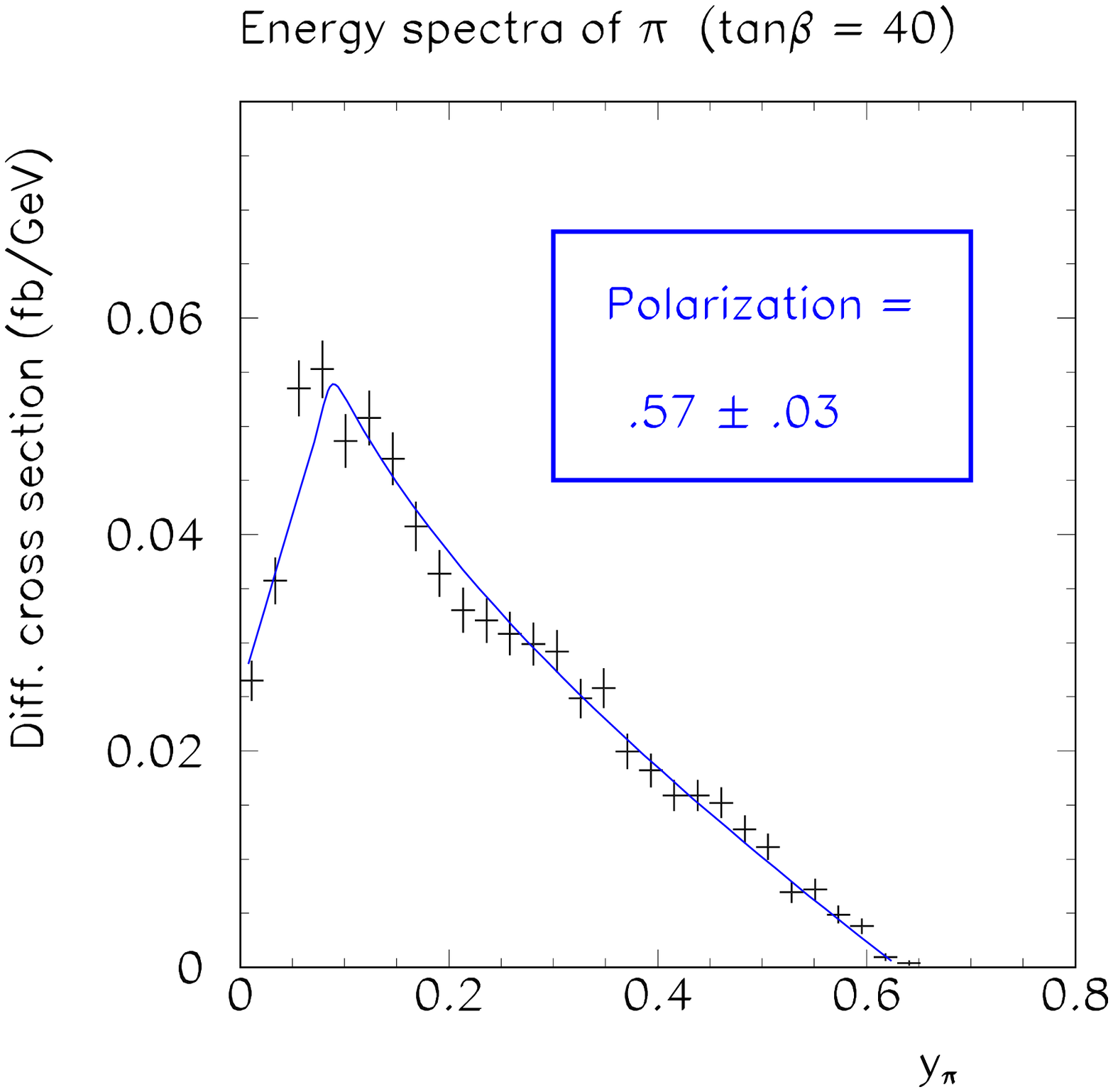}\hfill
\includegraphics[scale=0.45]{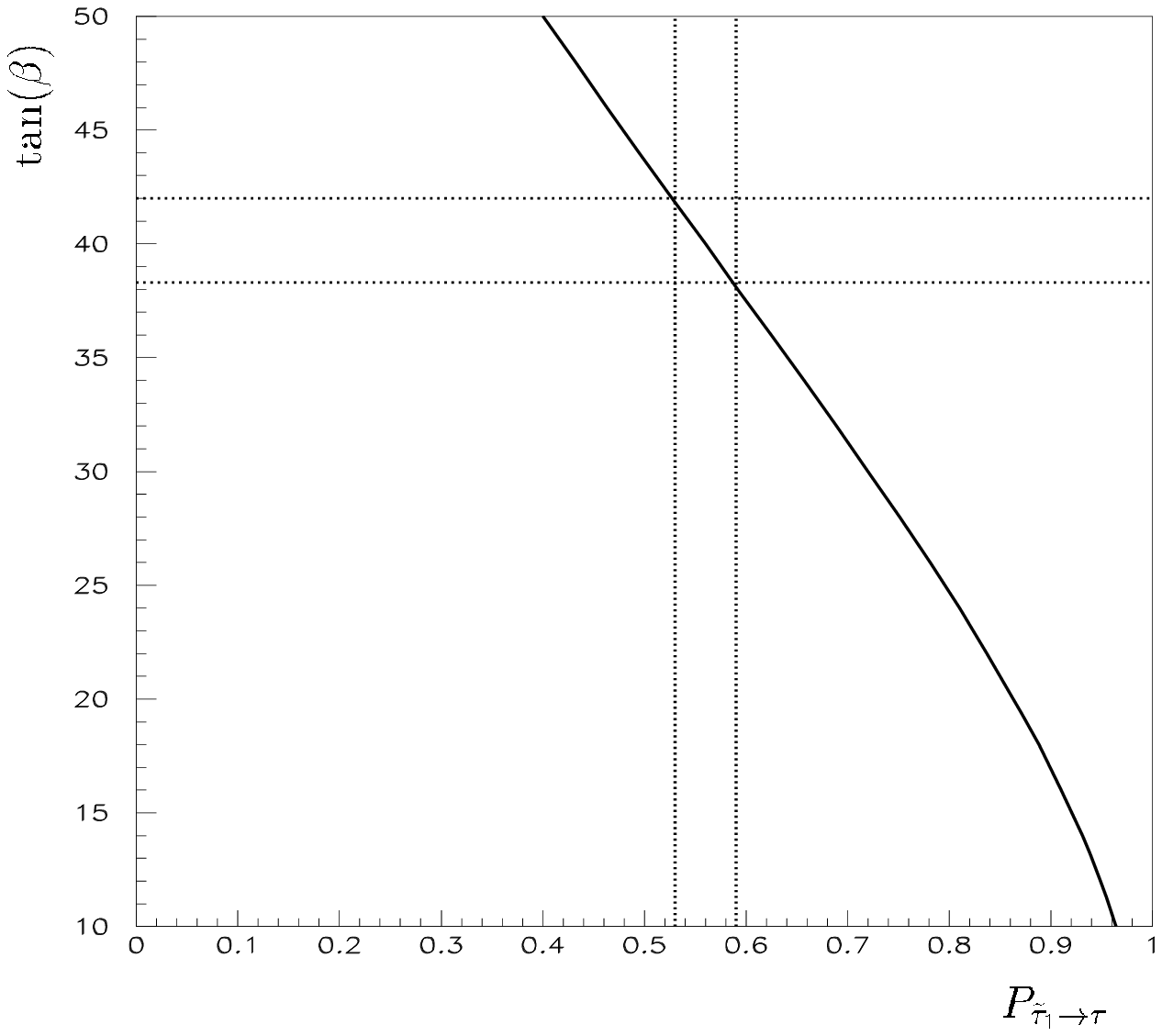}
\vspace*{-0.2cm}
\caption[]{\label{fig:lepton_tanb}
Left: $\pi$ energy spectrum fit ($y_\pi=2E_\pi/\sqrt{s}$) 
      determines $\tau$ polarization $P=0.57\pm 0.03$.
Right:\,resulting\,sensitivity\,$\tan\beta$$=$$40\pm2$\,for\,$\sqrt{s}=$500~GeV and $L=$500~fb$^{-1}$.
}
\vspace*{-0.7cm}
\end{figure}

The polarization of $\tau$ leptons from the decay of scalar taus
$\tilde\tau_1$ depends on $\tan\beta$.
The $\tau$ polarization 
can be measured from the shape of the energy spectrum of hadronic 
$\tau$ decays, for example in the reaction
$\rm \ee\rightarrow\tilde \tau^+_1\tilde\tau^-_1
        \rightarrow\tilde\tau^+_1\tau^-\tilde\chi^0_1
        \rightarrow \tilde\tau^+_1\nu_\tau\pi^-\tilde\chi^0_1$,
as illustrated in Fig.~\ref{fig:lepton_tanb}\cite{boos}.

\vspace*{-0.2cm}
\section{$\tan\beta$ Sensitivity from Higgs Boson Reactions}
\vspace*{-0.2cm}

\begin{figure}[htb]
\begin{minipage}{0.48\textwidth}
\includegraphics[scale=0.28,angle=90]{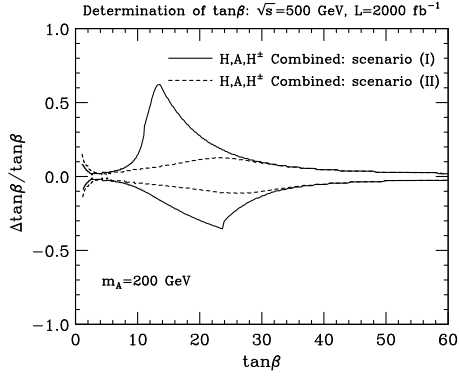}
\end{minipage}\hfill
\begin{minipage}{0.44\textwidth}
\caption[]{\label{fig:higgs_tanb}
Combined sensitivity on $\tan\beta$ from 
1) $\rm e^+e^- \rightarrow b\bar b \rightarrow b\bar b A\rightarrow
          b\bar b b\bar b$ rate;
2) $\rm e^+e^- \rightarrow HA \rightarrow b\bar b b\bar b$ rate;
3) H,A decay width;
4) $\rm e^+e^- \rightarrow H^+H^- \rightarrow t\bar b b\bar t$ rate;
5) $\rm H^+$ decay width.
Scenario I: no Supersymmetric particle decays, 
scenario II: Supersymmetric particle decays.
}
\end{minipage}
\vspace*{-0.4cm}
\end{figure}

Several Higgs boson reactions are particularly sensitive to 
$\tan\beta$ and excellent sensitivity on $\tan\beta$ 
from Higgs boson production rate and decay width measurements can be 
obtained (Fig.~\ref{fig:higgs_tanb})\cite{jack}.

\vspace*{-0.2cm}
\section{Theoretical and Experimental Precision}
\vspace*{-0.2cm}

As an example for the parameters of SPS-1a (mSUGRA) theoretical
precision\cite{kraml}, and expected experimental precision
a) from estimates of combined end-point and centre-of-mass scan 
methods\cite{grannis} 
and b) from detailed simulations\cite{martyn,nauenberg}
are summarized in Table~\ref{tab:precision}.
Some current theoretical uncertainties are much larger than the 
expected experimental precision.

\begin{table}[htp]
\vspace*{-0.1cm}
\tbl{Expected theoretical and experimental precision of 
     Supersymmetric particle masses for SPS-1a (mSUGRA) parameters.
     $\Delta_{\rm th}$:
     Gaussian error from
     Isajet 7.64, Softsusy 1.71, Spheno 2.0 and Suspect 2.101 
     studies.
     $ \Delta_{\rm exp}^{\rm a}$:
     estimated error for a 500~GeV $\rm e^+e^-$ LC 
     with ${\cal L}=1000$ fb$^{-1}$. 
     $\Delta_{\rm exp}^{\rm b}$:
     detailed simulations for a 400~GeV $\rm e^+e^-$ LC 
     with ${\cal L}=200$~fb$^{-1}$ ($\tilde \ell$ and $\tilde\chi^0_1$) and 
     for a 500~GeV $\rm e^+e^-$ LC 
     with ${\cal L}=500$ fb$^{-1}$ ($\tilde \nu_{\rm e}$).
}
{\footnotesize
\begin{tabular}{@{}c|ccccccccccc@{}}
\hline
[GeV]  & 
$m_{\tilde\chi_1^0}$ & $m_{\tilde\chi_2^0}$ &
$m_{\tilde\chi_1^\pm}$  & 
$m_{\tilde e_{\rm R}}$ & $m_{\tilde e_{\rm L}}$ &
$m_{\tilde \mu_{\rm R}}$ & $m_{\tilde \mu_{\rm L}}$ &
$m_{\tilde\tau_1}$  & $m_{\tilde\tau_2}$ &
$m_{\tilde \nu_{\rm e}}$ & $m_{\tilde\nu_{\rm\tau}}$ \\
\hline
$\Delta_{\rm th}$  & 0.3 & 1.5 & 1.2 & 0.6 & 1.1 & & & 1.1& 1.1 & 0.8 & 1.0 \\
$\Delta_{\rm exp}^{\rm a}$
                   & 0.07&0.12 & 0.18&0.02 & 0.2 &0.07 &0.51 &0.64& 1.1 & $\sim 1$ & ---\\
$\Delta_{\rm exp}^{\rm b}$
                   &0.10 &     &     &0.08 &     &0.17 &     &0.30&     & 0.7      &    \\
\hline
\end{tabular}\label{tab:precision}
}
\vspace*{-0.4cm}
\end{table}

\vspace*{-0.2cm}
\section{Synergy between the LHC and a LC}
\vspace*{-0.2cm}

\begin{figure}[htb]
\vspace*{-0.5cm}
\includegraphics[scale=0.4]{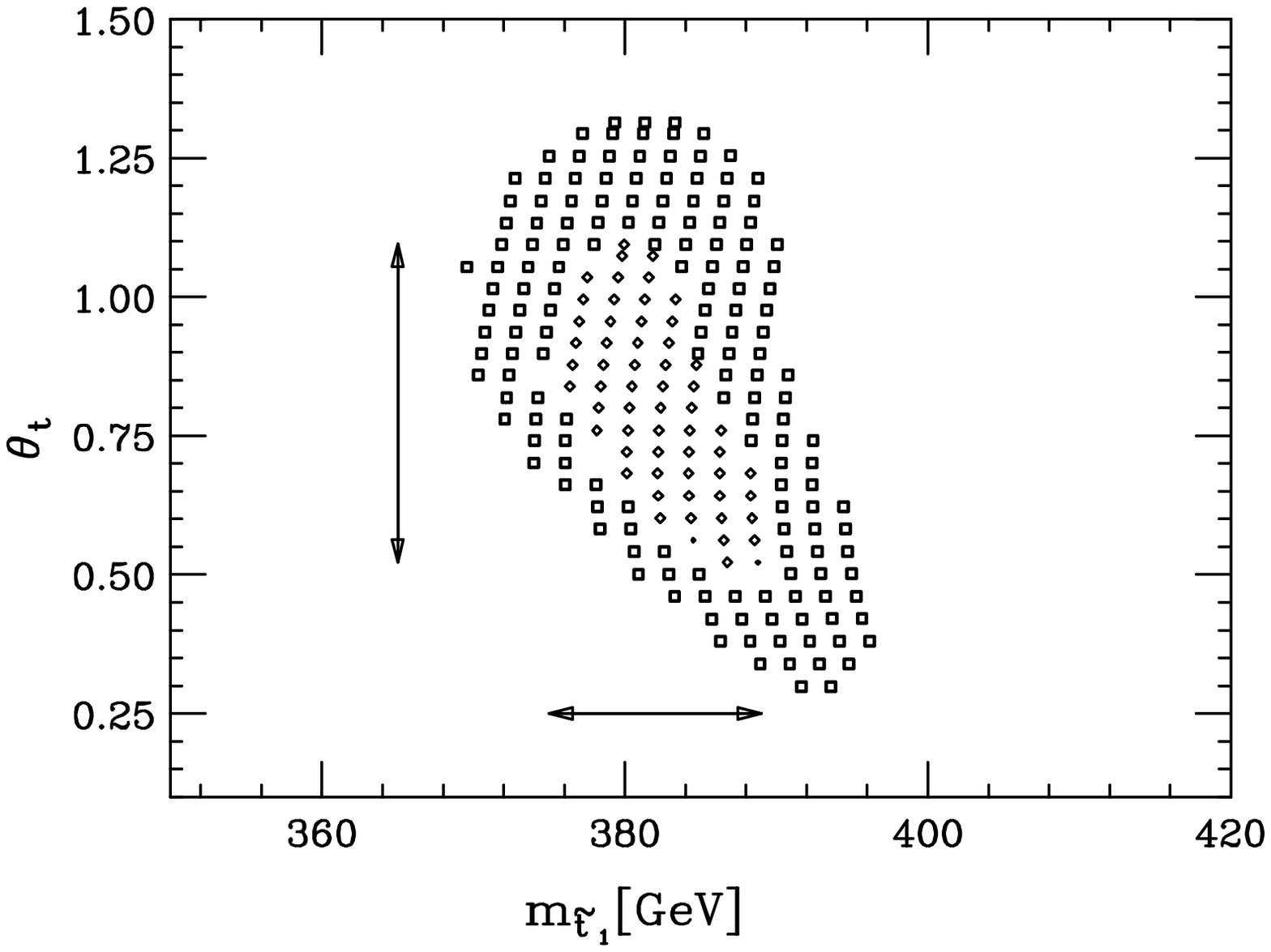}\hfill
\includegraphics[scale=0.27]{georg_mhAt02.cl.eps}
\vspace*{-0.2cm}
\caption[]{\label{fig:synergy}
Left: scalar top from LHC and light Supersymmetric particles from LC.
LC\&LHC give $\Delta m_{\tilde t_1}=\pm 7$~GeV and
$\Delta \theta_{\tilde t}=\pm 0.3$, but 
no mass and mixing sensitivity can be obtained from the LHC alone.
Right:
trilinear coupling precision $\Delta A_{\rm t} = \pm20$~GeV when
in addition to the LHC measurements (light-shaded band), 
precise top mass measurements from a LC (dark-shaded band) are used.
}
\vspace*{-0.5cm}
\end{figure}

For the benchmark parameters of SPS-1a all scalar leptons are in the 
kinematic reach of a 500 GeV LC and the scalar top quark masses are 
inaccessible at a LC,
but in reach of the LHC sensitivity.
Figure~\ref{fig:synergy} shows examples of the potential for combined LC 
and LHC studies\cite{lhclc_nojiri,lhclc_sven}.

\clearpage
\section{Distinguishing Supersymmetry Breaking Models}
\vspace*{-0.2cm}

At a future LC different Supersymmetry breaking models could be 
distinguished by their characteristic signatures.
For example in Anomaly Mediated Supersymmetry Breaking (AMSB)
models, which predict small
$\Delta m = m_{\tilde\chi^+_1} - m_{\tilde\chi^0_1}$ values,
the reaction $\ee\rightarrow\tilde\chi_1^+\tilde\chi_1^-(\gamma_{\rm ISR})$
has been studied\cite{hensel}.
In the decay mode $\tilde\chi^+_1 \rightarrow \tilde\chi^0_1\pi^+$ 
this analysis is based on tagging of initial-state
radiation photons $\gamma_{\rm ISR}$.
A fit of the $\pi$ spectrum 
$E_\pi \approx \Delta m$ gives $\Delta m = 0.413\pm0.017$~GeV
for $\sqrt{s}=500$~GeV and $L=500$~fb$^{-1}$.

Another example is the Gauge Mediated Symmetry
Breaking (GMSB) model with the reaction 
      $\rm e^-\gamma\rightarrow\tilde\chi^0_1\tilde e_{\rm R}^-
       \rightarrow \tilde e_{\rm R}^- e^- \tilde e_{\rm R}^-
       \rightarrow e^-e^-e^-\tilde G\tilde G$\cite{ghosal},
where a triple electron and missing energy signature is expected
in an $\rm e^-\gamma$ collider.
In this model the lightest Supersymmetric particle is the gravitino
$\tilde G$.

\vspace*{-0.2cm}
\section{$\gamma\gamma$ Collider and R-Parity Violation}
\vspace*{-0.2cm}

\begin{figure}[htb]
\vspace*{-0.3cm}
\begin{minipage}{0.53\textwidth}
\includegraphics[scale=0.4, bbllx=1pt,bblly=1pt,bburx=428pt,bbury=200pt,clip=]{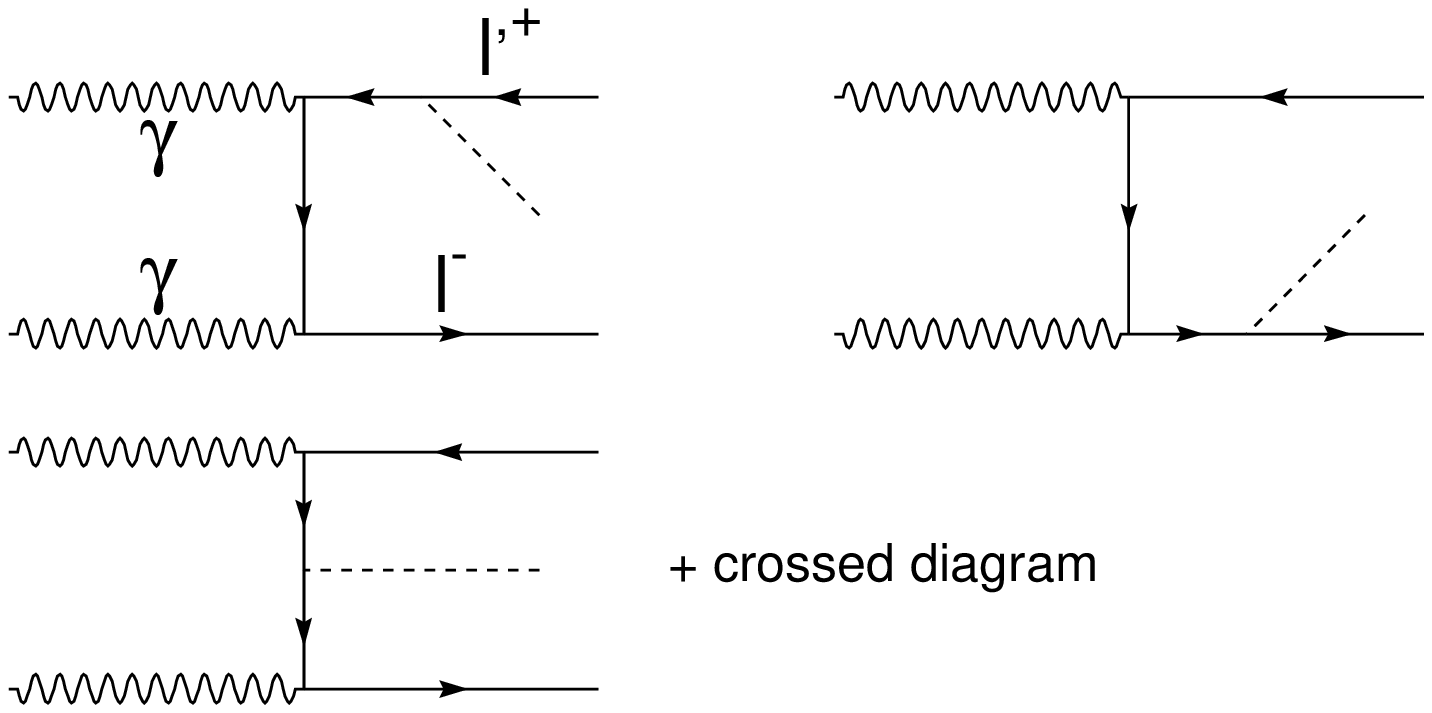}
\end{minipage} \hfill
\begin{minipage}{0.45\textwidth}
\includegraphics[scale=0.7, bbllx=210pt,bblly=312pt,bburx=423pt,bbury=527pt,clip=]{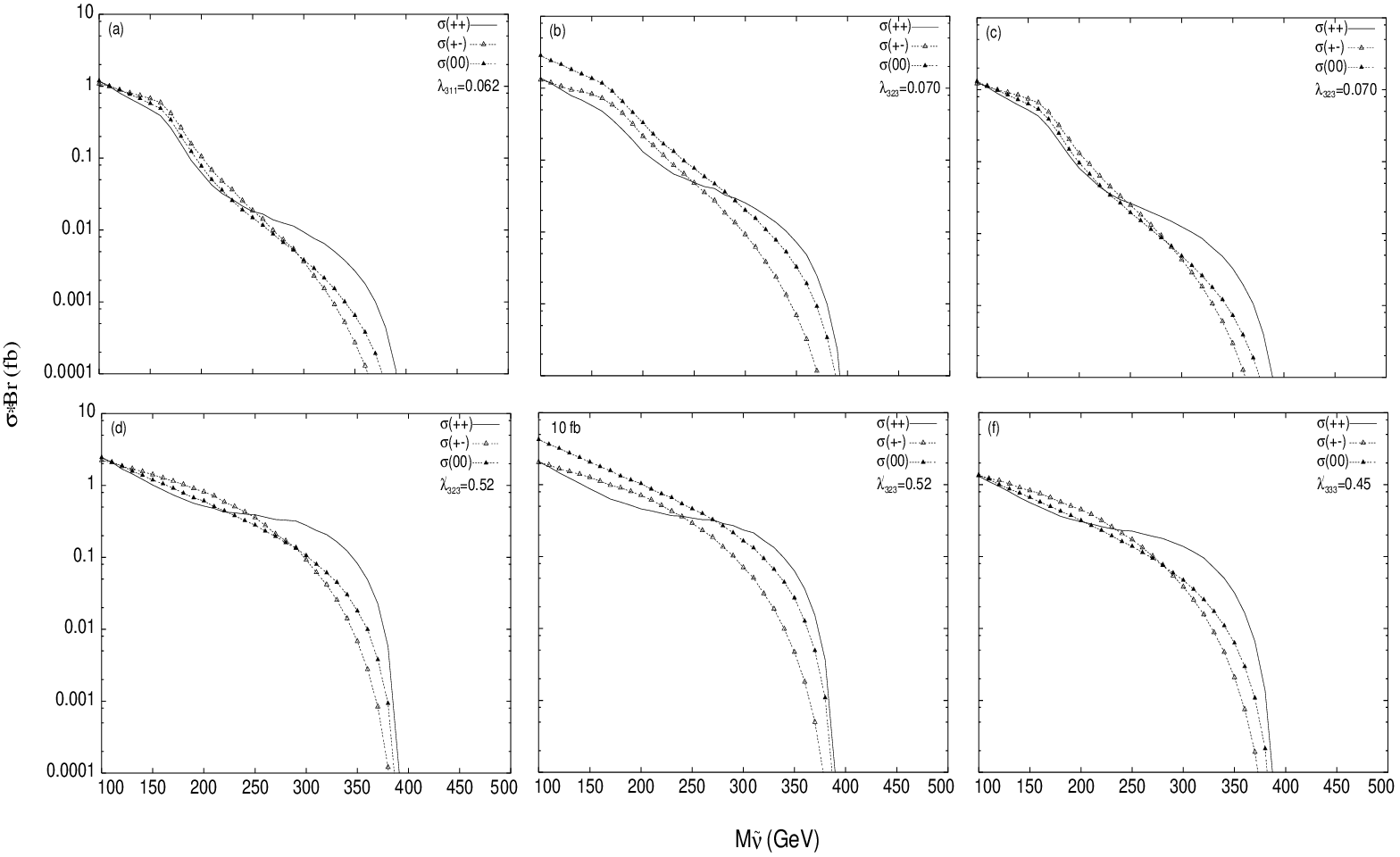}
\end{minipage}
\caption[]{\label{fig:gammagamma}
Left: $\rm \gamma\gamma\rightarrow f\bar f \tilde \nu$
      graphs, $\tilde \nu$ (dotted line).
Right: $\sigma\times BR( \tilde \nu \rightarrow f\bar f)$
       for different beam polarization states.
}
\vspace*{-0.6cm}
\end{figure}

The option in the LC project of photon collisions opens new fields
of research. For example, 
in a photon--photon collider, scalar neutrinos could be produced
via the reaction
$\rm \gamma\gamma\rightarrow f\bar f \tilde \nu\rightarrow f\bar f f\bar f$
which is almost free of Standard Model background in the
$\mu^+\mu^-\tau^+\tau^-$ final state\cite{moretti}.
The Feynman graphs and expected production cross sections are shown in 
Fig.~\ref{fig:gammagamma}.

\vspace*{-0.2cm}
\section{$\rm e^-e^-$ Collider}
\begin{figure}[htb]
\vspace*{-0.8cm}
\includegraphics[scale=0.3]{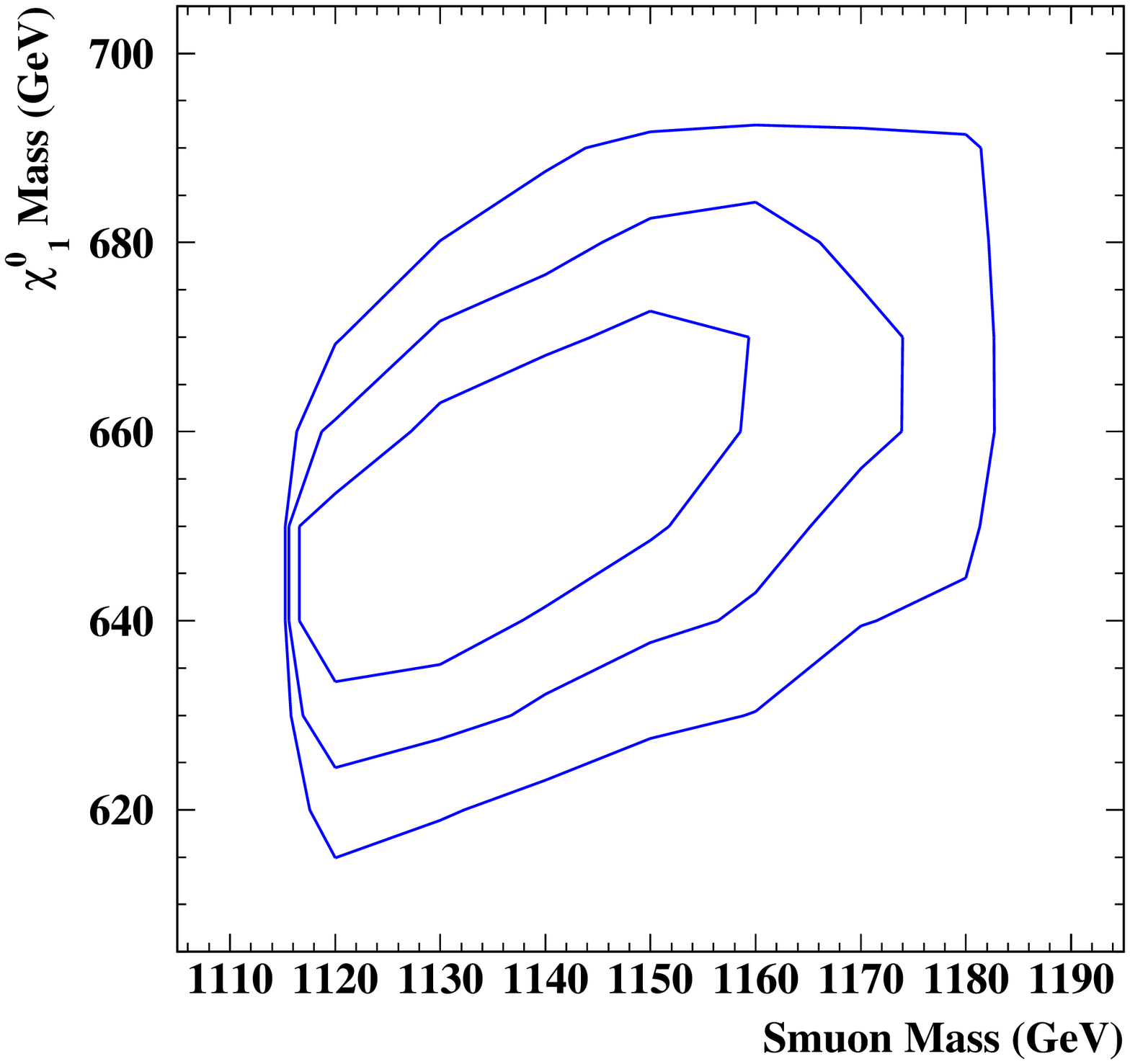}\hspace*{-0.5cm}
\includegraphics[scale=0.3]{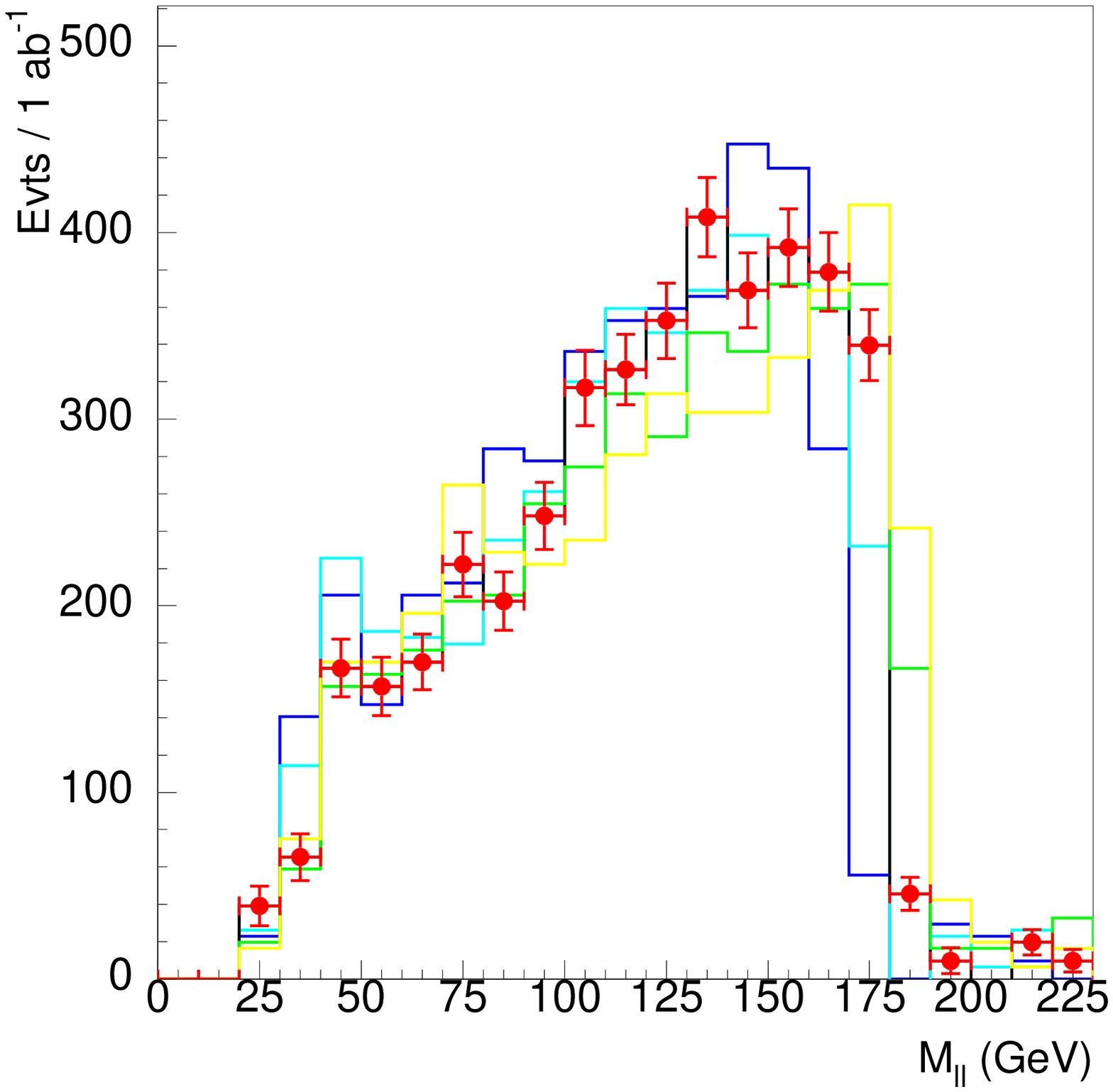}
\vspace*{-0.7cm}
\caption[]{\label{fig:clic}
Left:  reconstructed scalar muon 
       and neutralino $\tilde\chi^0_1$ masses.
Right: $\tilde\chi^0_2$ mass from the muon mass spectrum
       in one hemisphere 
       $\Delta m_{\tilde\chi_2^0} /  m_{\tilde\chi_2^0} = 1.7\%$.
}
\vspace*{-0.7cm}
\end{figure}

{
\vspace*{-0.2cm}

Another option in the LC project, $\rm e^-e^-$ collisions,
allows precision measurements beyond the sensitivity of an
$\rm e^+e^-$ LC.
Figure~\ref{fig:emem} shows as an example 
the production cross section for
$\rm e^-e^-\rightarrow \tilde e^-_{\rm R}\tilde e^-_{\rm R}
       \rightarrow e^- e^-\tilde\chi^0_1\tilde\chi^0_1$
and the precision of the mass determination of scalar electrons.
With a luminosity of only 10~fb$^{-1}$ the expected precision is 
about 20 MeV\cite{feng}.
 
\begin{figure}[htb]
\vspace*{-0.2cm}
\includegraphics[scale=0.36]{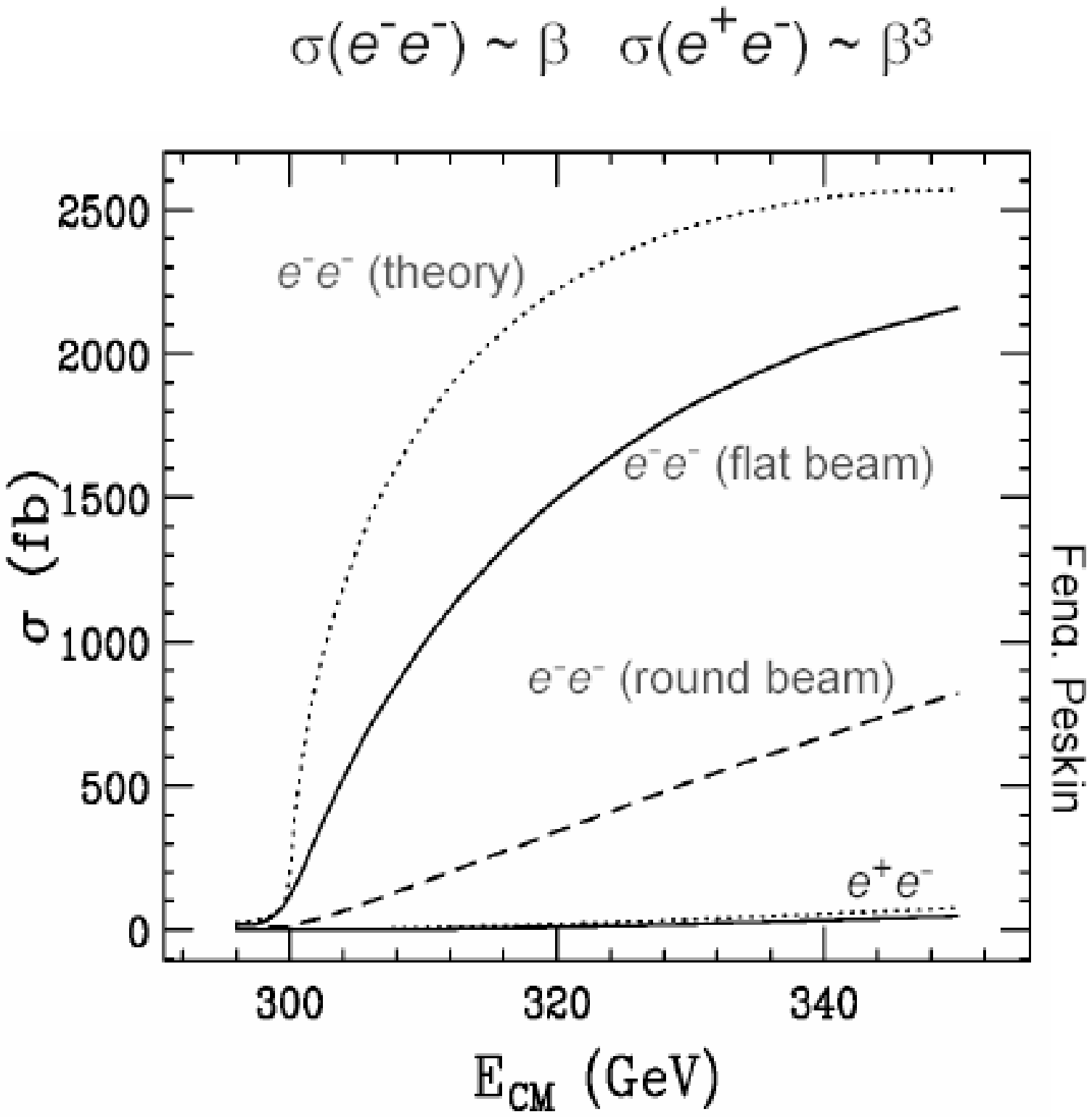}\hspace{-0.7cm}
\includegraphics[scale=0.36]{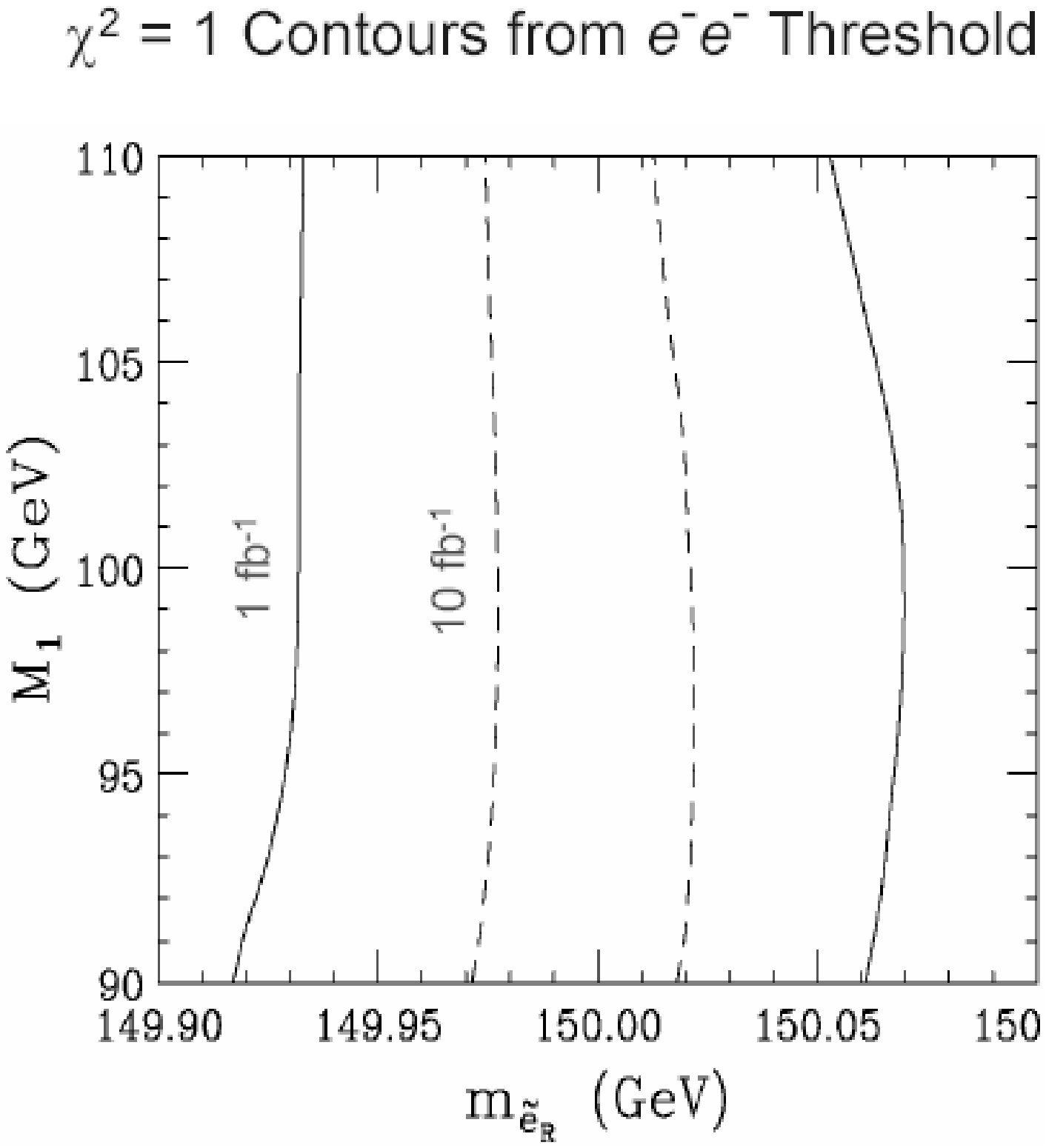}
\vspace*{-1cm}
\caption[]{\label{fig:emem}
Left:  $\rm e^-e^-\rightarrow \tilde e^-_{\rm R}\tilde e^-_{\rm R}
       \rightarrow e^- e^-\tilde\chi^0_1\tilde\chi^0_1$ 
production cross section.
Right: precision of the mass determination of scalar electrons
for $\sqrt{s}=300$~GeV and $P_{\rm e^-}=0.8$.
}
\vspace*{-0.2cm}
\end{figure}

\vspace*{-0.2cm}
\section{CLIC 3 TeV}
\vspace*{-0.2cm}

The CLIC project aims at higher centre-of-mass energies.
An example of the expected precision on the masses
of scalar muons, and neutralinos $\tilde\chi^0_1$ and $\tilde\chi^0_2$ 
from the reaction 
$\rm \ee\rightarrow \tilde\chi^0_2\tilde\chi^0_2 
 \rightarrow\mu^-\tilde \mu^+_{\rm R} \mu^+\tilde\mu^-_{\rm R}
 \rightarrow\mu^-\mu^+\tilde\chi^0_1 \mu^+\mu^-\tilde\chi^0_1$}
is given in Fig.~\ref{fig:clic}\cite{battaglia}.

\vspace*{-0.2cm}
\section{Conclusions}
\vspace*{-0.2cm}

Exploring Supersymmetry at a future linear collider is a 
very active field of research with many new ideas and new directions.
These studies also contribute to the detector design, for example with
c-quark tagging as a benchmark for vertex detectors.

After a first discovery at the Tevatron or the LHC and initial
precision measurements, the production and decay modes of many 
Supersymmetric particles will be measured with very high precision 
in the first phase of a LC.
The Linear Collider will probe the underlying production and decay mechanisms.
Detailed studies of several benchmark scenarios have been performed and 
Supersymmetry breaking models like mSUGRA, AMSM, or GMSB will be 
distinguished for a wide range of parameters.
Combined LC and LHC physics will expand the precision measurements and allow
for important consistency checks of the model.
The physics case for a future LC is established and the 
High-Energy-Physics community is 
ready to embark on the construction of the future global LC.

\vspace*{-0.3cm}
\section*{Acknowledgments}
\vspace*{-0.2cm}

I would like to thank the organizers of the SUSY03 conference for their 
kind hospitality, my colleagues from the Asian, European and 
US Supersymmetry working groups for help with the preparation of the
presentation, and 
Alex Finch,
Jan Kalinowski, 
Uli Martyn
and 
Hanna Nowak
for comments on the manuscript.


\end{document}